\documentclass[twocolumn,showpacs,preprintnumbers,amsmath,amssymb]{revtex4}

\usepackage{graphicx}
\usepackage{dcolumn}
\usepackage{bm}

\begin{document}

\title{Backward Raman compression of x-rays in metals and warm dense matters}

\author{S. Son$^1$\footnote[1]{Electronic address: {\sf seunghyeonson@gmail.com}}, S. Ku$^2$\footnote[2]{Electronic address: {\sf sku@cims.nyu.edu}},
and Sung Joon Moon$^3$\footnote[3] {Current address: Citigroup, 388 Greenwich St. New York, NY 10013}}
\affiliation{{$^1$18 Caleb Lane, Princeton, NJ 08540}\\
{$^2$Courant Institute of Mathematical Sciences, New York University, New York, NY 10012}\\
{$^3$PACM, Princeton University, Princeton, NJ 08544}}
\date{\today}

\begin{abstract}

Experimentally observed decay rate of the long wavelength Langmuir wave in metals
and dense plasmas is orders of magnitude larger than the prediction
of the prevalent Landau damping theory.
The discrepancy is explored, and the existence of a regime where the forward Raman
scattering is stable and the backward Raman scattering is unstable is examined.
The amplification of an x-ray pulse in this regime, via the backward Raman compression,
is computationally demonstrated, and the optimal pulse duration and intensity is
estimated.
\end{abstract}


\maketitle

Coherent intense x-rays would enable various
potential applications~\cite{fast3, fast4},
however, it is very difficult to compress an x-ray pulse, or even
in the UV regime.
Some progress has been made in this direction, based on the recent advances
in the areas of free electron laser~\cite{Free2} and the inertial
confinement fusion~\cite{Tabak}.
It is shown that the backward Raman scattering (BRS)~\cite{Fisch3,Fisch4},
where two light pulses and a Langmuir wave interact one another to exchange
the energy, is one promising approach to create an ultra short light pulse
(down to a few atto-seconds~\cite{Fisch,Fisch2}); a pulse gets compressed
and intensified via this laser-plasma interaction.
The BRS is successfully used to create intense pulses of the visible light
frequencies~\cite{Ping,Ping2}. It is natural to consider 
the same technique for the x-ray compression~\cite{Fisch,Fisch2}.
However, some physical processes in the regime where x-rays might be compressible
are considerably different from those in the visible light regime.
There are new physical processes to be considered, such as the Fermi degeneracy
and the electron quantum diffraction~\cite{sonprl,sonpla,IAW,sonlandau},  
to mention a few.

We consider one key aspect for the BRS in the x-ray compression regime,
i.e., the damping of the Langmuir wave.  
In an ideal plasma, the decay rate of the Langmuir wave increases rapidly
as the wavelength decreases.
It poses a concern for the BRS, as the plasmon from the forward Raman
scattering (FRS), having a lower wave vector than the BRS, depletes the pump.
There have been various attempts to suppress the FRS in the visible light
regime~\cite{Fisch5}.  
The situation is different in metals and the warm dense matters that we consider
for the x-ray compression.
The interaction of electrons via the inter-band transition (the Umklapp
process)~\cite{Sturm, Sturm2, Gibbons, Ku, Ku2, Pitarke, Thira} becomes the dominant plasmon decay
process, in contrast to the Landau damping in an ideal plasma.
As a consequence, the decay rate of the long wavelength plasmon is much higher
than the prediction of the prevalent dielectric function theory. 
In this paper, we show that a parameter regime where the BRS is unstable and
the FRS is stable does exist in metals and warm dense matters,
as the plasmon from the FRS strongly decays.
Here, we consider  metals in  room temperature, and show  that an x-ray pulse can be compressed in this regime.
We estimate an optimal pulse duration and intensity of the seed and pump pulses,
with which the inverse bremsstrahlung would be prevented.
We use an one-dimensional (1-D) simulation to demonstrate that a pump pulse
with a duration of hundreds of femto seconds could be compressed to a pulse
of a few or even sub-femto seconds.

The three wave interaction can be described by~\cite{McKinstrie}:
\begin{eqnarray}
\left( \frac{\partial }{\partial t} + v_1 \frac{\partial}{\partial x} + \nu_1\right)A_1  = -ic_1 A_2 A_3  \nonumber \mathrm{,}\\
\left( \frac{\partial }{\partial t} + v_2 \frac{\partial}{\partial x} + \nu_2\right)A_2  = -ic_2 A_1 A^*_3   \label{eq:2} \mathrm{,} \\
\left( \frac{\partial }{\partial t} + v_3 \frac{\partial}{\partial x} + \nu_3\right)A_3  = -ic_3 A_1 A^*_2  
\nonumber \mathrm{,}
\end{eqnarray}
where $A_i= eE_i/m_e\omega_ic$ is the amplitude
of the pump ($i=1$) and the seed ($i=2$) pulse,
$E_i$ is the electric field, $m_e$ is the electron mass,
$\omega_i$ is the pulse frequency, $c$ is the speed of the light,
$v_i$ is the group velocity of light (plasmon) scaled by $c$,
$\nu_i$ is the inverse bremsstrahlung rate,
$c_i = \omega_{\mathrm{pe}}^2/2\omega_i$, and
$k_i$ is the wave number of the pulses;
$A_3 = \tilde{n}_e/n_e$ is the Langmuir wave amplitude,
$\nu_3$ is the plasmon decay rate, and $c_3 = (cq)^2/2\omega_3$,
where $\omega_3 = \omega_{\mathrm{pe}}= \sqrt{4\pi n_e^2/m_e}$,
and $q$ is the Langmuir wave number.
In the BRS (FRS), the energy conservation reads
$\omega_1 = \omega_2 + \omega_{\mathrm{pe}}$ and the momentum conservation reads
$ q = k_1 + k_2 $ ($ q = k_1 - k_2 $).
The plasmon decay, $\nu_3$, and the inverse bremsstrahlung, $\nu_1$ and $\nu_2$,
are to be discussed later. 

The plasma density  response given by the third equation in Eq.(\ref{eq:2}) is derived from a fluid equation. However, it  can be still used in a degenerate or a partially degenerate plasma by the following reason. 
The plasma response is given as $\tilde{n}_e(q, \omega) =   \alpha(q, \omega_3) / (1 + 4 \pi e^2 /q^2  \alpha(q, \omega_3) )\phi_{\mathrm{pod}}$, where $\phi_{\mathrm{pod}} $ is the pondermotive potential from the beating of the seed and pump, and  $\epsilon = 1 +  (4 \pi e^2 /q^2) \alpha$ ($\alpha$) is the dielectric function (susceptibility), as an example of which,  in a degenrate plasma,  there is the well-known Lindhard function  in a zero temperature or a finite temperature \cite{Lindhard}.   It is shown by Capjack \cite{Capjack} that the fluid equation above describes reasonablly well the full kinetic plasma responce ($\tilde{n}_e $) for the classical dielectric function if $ 2 \nu_3 \cong  \mathrm{Im}(\epsilon) -i \mathrm{Re} (\epsilon)$.  Since  the theory to predict $\tilde{n}_e$ in a degenerate quantum plasma is not different from the one in classical plasmas,  the same fluid equation  can be used as long as an appropriate dielectric function has been calculated.   Here,  we choose the $\omega_3$ in such a way that $\mathrm{Re} (\epsilon)  =0$, 
then the damping, $\nu_3$, is purely real, which we will assume.

\begin{figure}
\scalebox{1.2}{
\includegraphics[width=0.775\columnwidth]{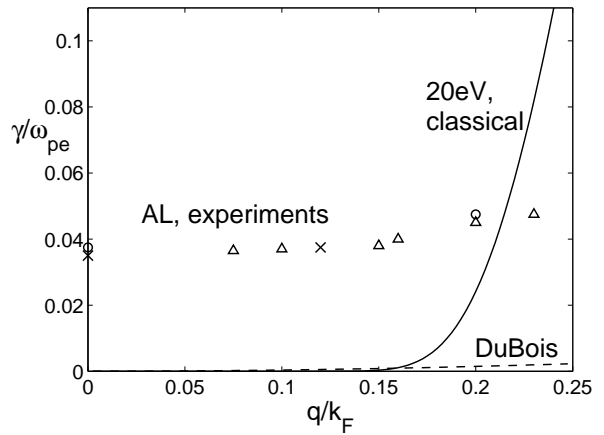}}
\caption{\label{fig:1} The damping rates of plasma of
$n_e = 1.8\times 10^{23} /\mathrm{cc}$;
the prediction from the classical dielectric function (solid line;
$T_e = 20 \mathrm{eV}$, chosen to be comparable to the Fermi energy) 
and from the DuBois' theory (dashed line; corresponding to $T_e=0$).
The experimental measurements for Al are adapted from Fig.~3 in
Ref.~\cite{Sturm2}, which are originally from Ref.~\cite{Zacharias}
(crosses), Ref.~\cite{Urner-Wille} (circles), and Ref.~\cite{Kloos}
(triangles).
}
\end{figure}

The Landau damping for the Langmuir wave, according to the dielectric function
theory, is given as 
\begin{equation}
\frac{\gamma}{\omega} = \frac{1}{2}\mathrm{Im}[\epsilon(q,\omega)] \mathrm{,}
\end{equation}
where $\gamma$ is the decay rate, $\epsilon$ is the dielectric function
satisfying $\mathrm{Re}[ \epsilon ] = 0$, and $q$ is the wave vector.
The Landau damping rate in the free-electron plasma can be computed by
using either the classical dielectric function or
the degenerate Lindhard dielectric function.
The damping rates from the experimental measurements and different theories
are compared in Fig.~\ref{fig:1}, in a range of $q$.
The damping rate from the Lindhard dielectric function~\cite{Lindhard}
is essentially zero in this range of $q$, which is not visible in this scale.
We note that DuBois' theory~\cite{DuBois}, accounting for the dynamical
correlation, predicts $\gamma$ is proportional to $(q/k_F)^2$
in the long wavelength limit (as $q$ goes to zero), where $k_F$ is the Fermi wave vector.
While all these theories predict the damping rate to vanish in the long wavelength
limit in a way or another, there are experimental evidences that the rates are finite in this limit. 
For example, experimental measurements in the $(1,0,0)$ direction of Al
exhibit  finite damping rates as shown in Fig.~\ref{fig:1}.
The experimentally observed plasmon decay rates do not vary much for $q < 0.5k_F$,
where the inter-band transition is the dominant decay mechanism.
It rapidly increases for $q > 0.5k_F$ (not shown in Fig.~\ref{fig:1};
see Fig.~3 in Ref.~\cite{Sturm2}), where the Landau damping becomes dominant.
We express the experimental decay rate of the plasmon for $q < 0.5 k_F$ as 
\begin{equation}
\nu_3(q) = \eta(q) \omega_{\mathrm{pe}}  \label{eq:lang} \mathrm{,}
\end{equation}
where $\eta(q) = \eta_0 + d\eta/dq^2 (q^2/k^2_F)$. 
For a typical metal, $ 0.02 < \eta_0 < 0.2$ and $ d\eta/dq^2 \cong a \eta_0 $,
where $ 2 < a < 10$~\cite{Gibbons}.
This experimental damping rate is successfully modeled by  Sturm~\cite{Sturm2},
who takes into account of the Umklapp process of electrons in the presence
of the spatially periodic ions. 

The inverse bremsstrahlung ($\nu_1$ and $\nu_2$) in metals is given as~\cite{son2},
\begin{equation}
\nu_i(\omega, E_i) = 4 \pi n_i Z_i^2  \frac{e^4}{m_e^2 v_F^3} \frac{\omega_{pe}^2}{\omega_i^2} \frac{F(\alpha_i)}{ s_i^2 \kappa_i^{1/2}}\mathrm{,}
\label{eq:brm}
\end{equation}  
where
$ \kappa_i =  \hbar \omega_i / 2 m_e v_F^2 < 1 $, $v_F$ is the Fermi energy,
$s_i =  e E_i / m_e \omega_i v_F$,  
$ \alpha_i^2 = 2e^2E_i^2/m\hbar \omega_i^3 $, and 
$ F(\alpha) $ can be approximated as $\alpha^2/6 $ for $\alpha < 2$
and $(2/\pi\alpha)\log(2\alpha)^2$ otherwise.
In Eq.~(\ref{eq:brm}), it is assumed that $\hbar \omega > T_e$ and $\hbar \omega > E_F$,
where $T_e$ and $E_F$ is the electron temperature and the Fermi energy, respectively.  
For $\alpha < 2$, the above equation can be simplified as 
\begin{equation}
 \nu_i = 2.9 \times 10^{15} Z_i \kappa_i^{-3/2} \frac{\omega_{\mathrm{pe}}^2}{\omega_i^2}~~~\sec^{-1} \label{eq:brem} \mathrm{.}
\end{equation}
When $1/\nu_i $  is shorter than the pulse duration,
the pump pulse would get heavily damped before having a chance to compress
the seed pulse.
Note that $\nu_1$ is proportional to $(\omega_{\mathrm{pe}}/\omega_1)^2$.

From Eqs.~(\ref{eq:2}) and (\ref{eq:lang}), the BRS growth rate,
assuming that the pump intensity is large enough to make it unstable,
is roughly estimated as  
\begin{equation}
g_B =  \frac{A_1^2}{\eta(q)} \frac{\omega_2 }{\omega_{\mathrm{pe}}} \label{eq:g} \omega_{\mathrm{pe}}   \mathrm{.}
\end{equation} 
The larger $g_B$ is, the stronger the BRS is.
It can be seen from Eqs.~(\ref{eq:brem}) and (\ref{eq:g}) that
$ \omega_1 /\omega_{\mathrm{pe}} \gg 1$
is desirable, as this condition would suppress the inverse bremsstrahlung,
while keeping the BRS growth rate high.
In order to maximize the efficiency, we choose $\omega_i/\omega_{\mathrm{pe}}$
so that the wave vector $(\omega_1+\omega_2)/c $ does not exceed $0.25k_F$,
beyond which the linear Landau damping is important.
In various metals, the desirable $\omega_i/ \omega_{\mathrm{pe}}$ is  
\begin{equation}
10<\frac{\omega_1}{\omega_{\mathrm{pe}}}<30 \mathrm{,} \label{eq:freq}
\end{equation}
which corresponds to the photon energy in AL between 150 eV and 450 eV.

We obtain the possible operating regime for the Raman compression
using Eqs.~(\ref{eq:2}), (\ref{eq:lang}), (\ref{eq:brem}) and (\ref{eq:freq}).  
The stability condition for the FRS, obtained from Eq.~(\ref{eq:2}), is given to
be $ c_2c_3|A_1|^2 <  \nu_2 \nu_3 $, where $\nu_2$ and $\nu_3$ can be found in
Eqs.~(\ref{eq:lang}) and (\ref{eq:brem}), and $c_3= \omega_{\mathrm{pe}}/2$.
This condition can be written as $|A_1| > |A_{1F}|$, where 
\begin{equation}
 |A_{1F}|^2 = \frac{4 \nu_2(q_F) \nu_3}{\omega_{\mathrm{pe}}^2} \frac{\omega_2}{\omega_{\mathrm{pe}}}
 \label{eq:for} \mathrm{,}
\end{equation} 
and $q_F = \omega_{\mathrm{pe}}/c$.  
The same condition for the BRS is estimated to be $ |A_1| < |A_{1B}|$,
using $c_3 = (\omega_1 +\omega_2)^2/2\omega_{\mathrm{pe}}$, where 
\begin{equation}
 |A_{1B}|^2 = \frac{\nu_2(q_B) \nu_3}{\omega_{\mathrm{pe}}^2} \frac{4\omega_{\mathrm{pe}} \omega_2}{(\omega_2 +\omega_1)^2}
 \label{eq:back} \mathrm{,}
\end{equation} 
and $q_B = (\omega_1 +\omega_2)/c$.  
$|A_{1B}|^2$ is smaller than $ |A_{1F}|^2$ by a factor of
$(\omega_1 +\omega_2)^2/\omega_{\mathrm{pe}}^2 (\eta(q_F)/\eta(q_B))$,
which is considerably large since $\eta(q_F)/\eta(q_B) \cong 1$. 

A regime, where the BRS is unstable
and the FRS is stable, exists due to the strong inverse bremsstrahlung and
a rather constant damping of the Langmuir wave in a range of $q$. 
 While these boundary is estimated assuming the maximum plasma wave amplitude  using the
large damping limit,  this is useful for an estimation. Furthermore, in the linear regime where the pump is uniform and the seed pulse is small so that the linear theory is valid,  the above boundary is exact. 
Even when the FRS is unstable, it would not be detrimental unless
the background noise plasmon grows to such an extent that depletes the pump.
The growth rate of the FRS, assuming unstable, is given as 
\begin{equation} 
g_F = 0.25 \times  \frac{1}{ \eta} \frac{\omega_{\mathrm{pe}}}{\omega_2} |A_1|^2 \omega_{\mathrm{pe}}  \label{eq:growth} \mathrm{.} 
\end{equation}
 Unless $g_F \tau_1 > 10$ , where $\tau_1$ is the duration of the pump, the FRS instability  can be ignored. 
In AL, $\eta$ is roughly 0.03.
Even for $\omega_{\mathrm{pe}}\tau_1=10^4$,
the pump pulse intensity has to be unrealistically large for the FRS depletion to occur.
In other words,  even though the FRS is absolutely unstable (linearly unstable), 
the FRS/RBS is convectively stable.

Now a detailed estimation using the parameter for AL is in order, assuming $Z_i =3$, $\eta = 0.03$,
$\hbar \omega_{\mathrm{pe}} = 15.3 \mathrm{eV}$, and
$E_F = 11.7 \mathrm{eV}$, with which the electron density $n_e$ is
$1.8 \times 10^{23} / \mathrm{cc}$ and $\omega_{\mathrm{pe}} = 2.4 \times 10^{16} \sec^{-1} $.
Consider the case when $\omega_1/\omega_{\mathrm{pe}} = 10 $
(and 20).
The inverse bremsstrahlung for the pump and the seed is roughly given as
$\nu_{1,2} = 7 \times 10^{-4} \omega_{\mathrm{pe}}$
($\nu_{1,2} = 5 \times 10^{-5} \omega_{\mathrm{pe}}$).
The threshold for the FRS in Eq.~(\ref{eq:for}) is estimated to be
$ |A_{1F}|^2 = 6.72\times 10^{-4}$ ($|A_{1F}|^2 = 1.19\times 10^{-4}$),
which corresponds to $1.27 \times 10^{19}~\mathrm{W}/\mathrm{cm}^2 $
($8.98 \times 10^{18} \mathrm{W}/\mathrm{cm}^2 $).
On the other hand, the estimated threshold for the BRS in Eq.~(\ref{eq:back}) is
$|A_{1B}|^2 = 1.68 \times 10^{-7}$, or $ 3.1 \times 10^{16}~\mathrm{W} /\mathrm{cm}^2 $
($|A_{1B}|^2 = 3.3 \times 10^{-7}$, or $ 6.1\times 10^{16} \mathrm{W} /\mathrm{cm}^2$). 

\begin{figure}
\scalebox{1.0}{
\includegraphics[width=0.895\columnwidth]{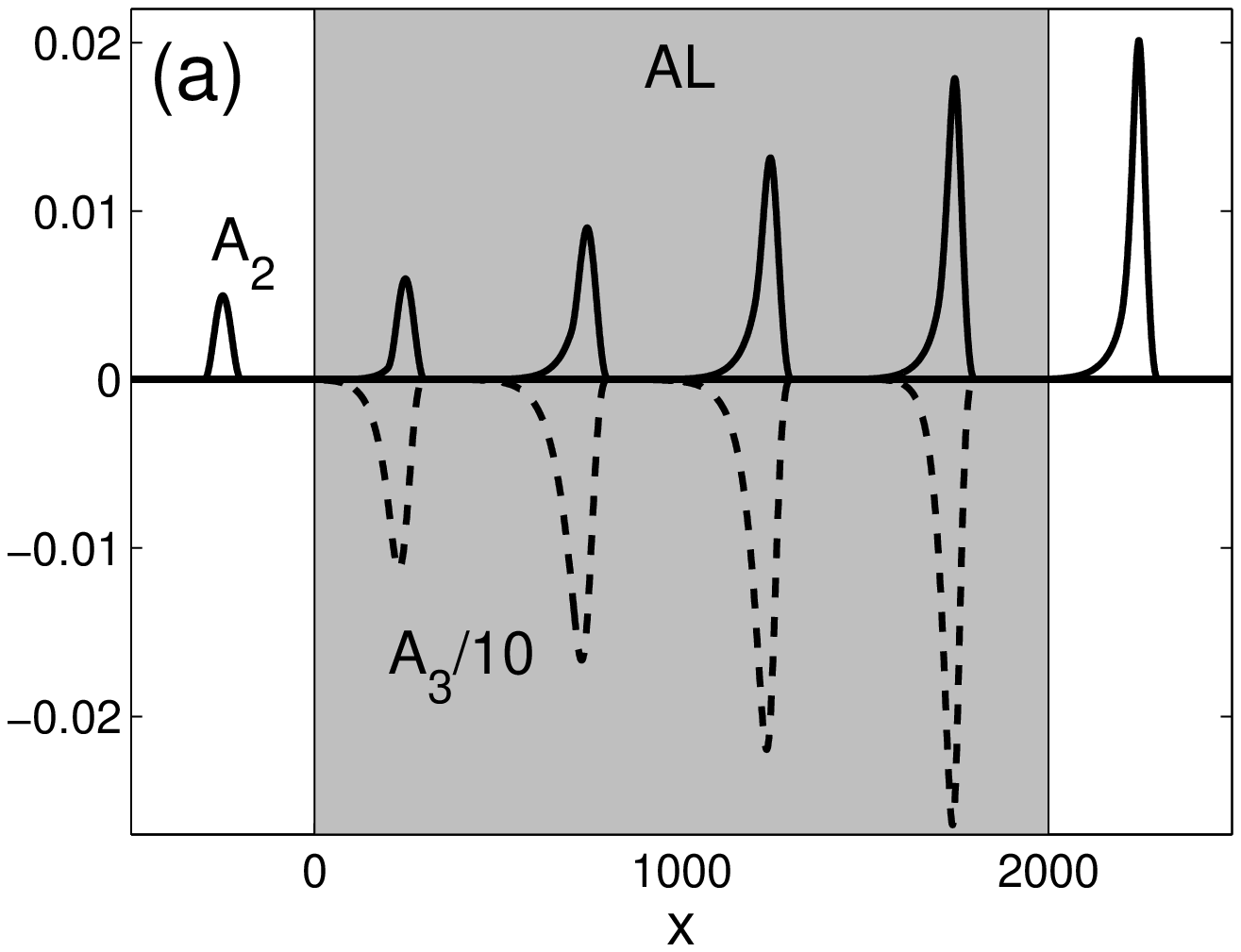}}
\scalebox{1.0}{
\includegraphics[width=0.895\columnwidth]{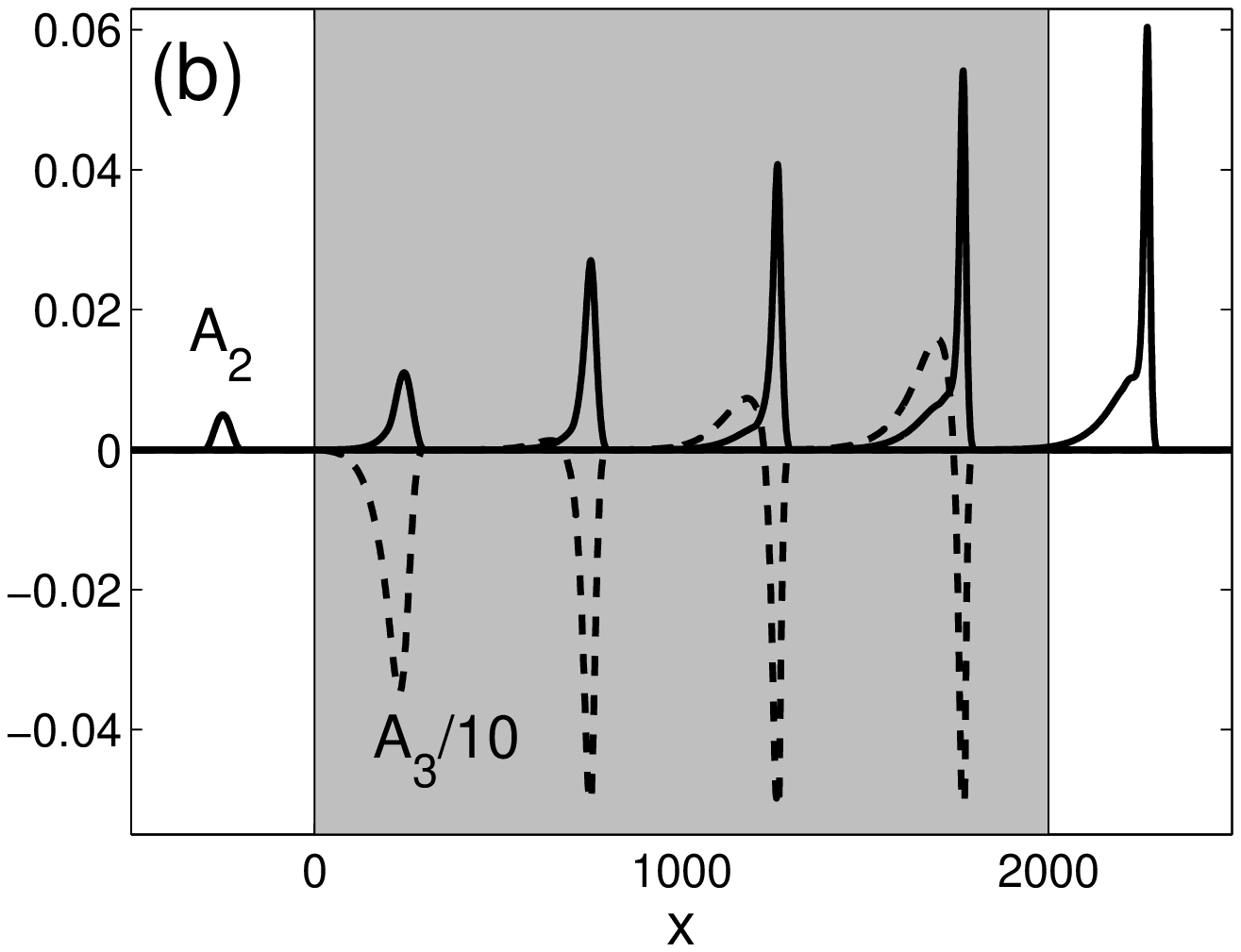}}
\caption{\label{fig:2} Time series of the seed pulse,
$A_2 = eE_2/m_e\omega_2 c$,
and the Langmuir wave, $ A_3 = \tilde{n} / \bar{n} $,
traveling to the right through AL (gray area, $ 0\le x \le2000$),
where $x$ is normalized by $ \omega_{\mathrm{pe}}/c $.
Each pulse is separated by 5000/$\omega_{\mathrm{pe}}$, and
the duration of the pump pulse is $4000/\omega_{\mathrm{pe}}$ in both cases.  
See the text for details.
}
\end{figure}

We numerically integrate the 1-D model, Eq.~(\ref{eq:2}), to demonstrate
the compression of an x-ray pulse in AL (Fig.~\ref{fig:2}).
The pump pulse travels from the left to the right as does the seed pulse.
The initial intensity of the pump is chosen to be
$ 2.5 \times 10^{17}~\mathrm{W}/\mathrm{cm}^2 $ in Fig.~\ref{fig:2} (a)
and  $10^{18}~\mathrm{W}/\mathrm{cm}^2 $ for Fig.~\ref{fig:2} (b).
The seed is $10^{18}~\mathrm{W}/\mathrm{cm}^2 $ and the pump frequency is
$\omega_1 /\omega_{\mathrm{pe}} = 15$ in both cases.
The seed pulse extracts the energy from the pump and
the intensity gets amplified by a factor of about 15 (Fig.~\ref{fig:2} (a))
and 140 (Fig.~\ref{fig:2} (b)).
When the nonlinear pump depletion is negligible, $\tau_B=1/g_B$ in Eq.~(\ref{eq:g})
is the growth time of the seed pulse, which roughly corresponds to the width
of the compressed pulse.
When the nonlinear pump depletion becomes important, the pump depletion time might be
shorter than the linear growth time $\tau_B$ by a factor of $|A_2 / A_1|^2$.
Since the pump feeds the energy into a small portion of the seed pulse
in the nonlinear depletion regime, the actual compressed pulse width would be much
narrower than the prediction of our linear analysis in Eq.~(\ref{eq:g}).
In Fig.~\ref{fig:2} (a),
the maximum of the Langmuir wave intensity $|A_3|$ during the compression
remains less than about 0.3.

For an intense Langmuir wave, the nonlinear cascading could be an important saturation
factor~\cite{Drake}. It is estimated that the nonlinear cascading is not important in the compression time scale ($\tau_c = 10/ \omega_{\mathrm{pe}}$). 
In addition, the damping of the Langmuir wave is strong as shown here,  
and the damping of the ion acoustic wave is strong
due to the Umklapp process.  
Consequently, the minimum threshold for the nonlinear cascading~\cite{Drake}
would be high, so it becomes of less concern in dense plasmas.

Now we consider the wave breaking.
The maximum strength of the Langmuir wave is
$ |A_3|_{\mathrm{max}} = |A_1| |A_2| (\omega_1 +\omega_2)^2/\omega_{\mathrm{pe}}^2 \eta(q)$
(the third equation in Eq.~(\ref{eq:2})).
The theoretically achievable value of $|A_3|_{\mathrm{max}}$ is less than the unity,
which is called the wave breaking.
Before $|A_3|$ reaches the wave breaking, various saturation mechanisms come into play,
including the frequency detuning and the soliton formation~\cite{Rose, Oneil}.
These mechanisms make the linear theory presented here become insufficient.
In the simulation results shown in Fig.~\ref{fig:2}, the Langmuir wave frequency
is higher
than $E_F/\hbar $ and $k \cong 0.2 k_F$, and the phase velocity of the plasmon
is much higher than the Fermi velocity.
We estimate that, until $|A_3|$ reaches $\sim~0.5$, the frequency detuning due to
the trapped particles  is negligible and the above linear theory is valid,
which is the case in Fig.~\ref{fig:2} (a).
When $|A_3|_{\mathrm{max}}$ exceeds $\sim$ 0.5, the currently available theories~\cite{Rose}
need to be extended by taking into account of the electron degeneracy and
the quantum diffraction effects~\cite{sonprl, sonpla, sonlandau}. 
In Fig.~\ref{fig:2} (b), we assume that  the $|A_3|$ quickly decays when $|A_3| > 0.5$ (due to the wave breaking, the frequency detuning, and the nonlinear cascading).



The inverse bremsstrahlung would be of less concern as can be seen from Eq.~(\ref{eq:brm}),
and it would be further diminished by the electron degeneracy~\cite{sonprl,sonpla}.
As the pump travels inside the medium, it heats electrons via inverse bremsstrahlung especially on the back of the pump.  As the electron temperature increases, the inverse bremsstrahlung decreases faster. 
While a high electron temperature would certainly increase the Landau damping
and the partial wave breaking arising from the trapped particles, it is
suggested that the damping and the wave breaking would be weak in
dense plasmas~\cite{sonlandau}.  
 More concern for the heating of electrons would be the detuning effect of the plasmon from the change of the plasmon disperson relation, and the reduction of the plasmon damping.  The reduction of the plasmon damping would make the FRS more unstable, but our preliminary estimation suggests that this is still containable.  The more detailed estimation should wait until the thoory of the plasmon damping for the partially degenerate plasma is available, which is in progress.

To summarize, it is shown that the FRS is less harmful for the Raman compression
in metals and warm dense matters than in the gas discharge plasmas or ideal plasmas,
as a consequence of the enhanced decay of the long wavelength Langmuir waves.
The optimal frequency and the pulse duration for the compression is estimated
in our analysis.
Our 1-D simulation based on the linear analysis demonstrates that a pump pulse
with the duration of
$10^3/\omega_{\mathrm{pe}} -10^4/\omega_{\mathrm{pe}}$ and the intensity comparable to
$ 10^{18} \mathrm{W} / \mathrm{cm}^2 $ is compressed to the one of
$1/\omega_{\mathrm{pe}} -10/\omega_{\mathrm{pe}}$, with the intensity of
$10^{19} \mathrm{W} / \mathrm{cm}^2 - 10^{21} \mathrm{W} / \mathrm{cm}^2$. 

As suggested above, it is crucial to understand the plasmon damping rate
in the presence of the relic lattice structure.
The plasmon decay in metals strongly depends on the direction relative
to the lattice structure. 
As the electron temperature increases, the electron screening of the ions
would get diminished, which enhances the plasmon damping~\cite{Sturm2}.
As the electron heats up further, the ions would finally lose
their lattice structure.
A larger value of $\eta(q)$ in Eq.~(\ref{eq:lang}) would be preferred for a strong pump,
while a small value of $\eta(q)$ is better for a weak pump.  
The damping rate in the warm dense matters might be accessible by the measurement
of the electron energy loss in the thin heated foil experiments~\cite{Gibbons}. 
The effect of the phase transition on the plasmon decay is theoretically challenging
to understand, yet it is important because of its relevance to the BRS x-ray compression. 

The authors would like thank Prof.~Fisch for useful discussion and advice. 

\bibliography{backward1}

\begin{thebibliography}{33}
\expandafter\ifx\csname natexlab\endcsname\relax\def\natexlab#1{#1}\fi
\expandafter\ifx\csname bibnamefont\endcsname\relax
  \def\bibnamefont#1{#1}\fi
\expandafter\ifx\csname bibfnamefont\endcsname\relax
  \def\bibfnamefont#1{#1}\fi
\expandafter\ifx\csname citenamefont\endcsname\relax
  \def\citenamefont#1{#1}\fi
\expandafter\ifx\csname url\endcsname\relax
  \def\url#1{\texttt{#1}}\fi
\expandafter\ifx\csname urlprefix\endcsname\relax\def\urlprefix{URL }\fi
\providecommand{\bibinfo}[2]{#2}
\providecommand{\eprint}[2][]{\url{#2}}

\bibitem[{\citenamefont{Chapman}(2006)}]{fast3}
\bibinfo{author}{\bibfnamefont{H.~N.} \bibnamefont{Chapman}},
  \bibinfo{journal}{Nature Physics} \textbf{\bibinfo{volume}{2}},
  \bibinfo{pages}{839} (\bibinfo{year}{2006}).

\bibitem[{\citenamefont{Hentschel et~al.}(2001)\citenamefont{Hentschel,
  Kienberger, Spielmann, Reider, Milosevic, Brabec, and Corkum}}]{fast4}
\bibinfo{author}{\bibfnamefont{M.}~\bibnamefont{Hentschel}},
  \bibinfo{author}{\bibfnamefont{R.}~\bibnamefont{Kienberger}},
  \bibinfo{author}{\bibfnamefont{C.}~\bibnamefont{Spielmann}},
  \bibinfo{author}{\bibfnamefont{G.~A.} \bibnamefont{Reider}},
  \bibinfo{author}{\bibfnamefont{N.}~\bibnamefont{Milosevic}},
  \bibinfo{author}{\bibfnamefont{T.}~\bibnamefont{Brabec}}, \bibnamefont{and}
  \bibinfo{author}{\bibfnamefont{P.}~\bibnamefont{Corkum}},
  \bibinfo{journal}{Nature} \textbf{\bibinfo{volume}{414}},
  \bibinfo{pages}{509} (\bibinfo{year}{2001}).

\bibitem[{\citenamefont{Emma et~al.}(2004)\citenamefont{Emma, Bane, Cornacchia,
  Huang, Schlarb, Stupakov, and Walz}}]{Free2}
\bibinfo{author}{\bibfnamefont{P.}~\bibnamefont{Emma}},
  \bibinfo{author}{\bibfnamefont{K.}~\bibnamefont{Bane}},
  \bibinfo{author}{\bibfnamefont{M.}~\bibnamefont{Cornacchia}},
  \bibinfo{author}{\bibfnamefont{Z.}~\bibnamefont{Huang}},
  \bibinfo{author}{\bibfnamefont{H.}~\bibnamefont{Schlarb}},
  \bibinfo{author}{\bibfnamefont{G.}~\bibnamefont{Stupakov}}, \bibnamefont{and}
  \bibinfo{author}{\bibfnamefont{D.}~\bibnamefont{Walz}},
  \bibinfo{journal}{Phys. Rev. Lett.} \textbf{\bibinfo{volume}{92}},
  \bibinfo{pages}{074801} (\bibinfo{year}{2004}).

\bibitem[{\citenamefont{Tabak et~al.}(1994)\citenamefont{Tabak, Hammer,
  Glinsky, Kruerand, Wilks, Woodworth, Campbell, and Perry}}]{Tabak}
\bibinfo{author}{\bibfnamefont{M.}~\bibnamefont{Tabak}},
  \bibinfo{author}{\bibfnamefont{J.}~\bibnamefont{Hammer}},
  \bibinfo{author}{\bibfnamefont{M.~E.} \bibnamefont{Glinsky}},
  \bibinfo{author}{\bibfnamefont{W.~L.} \bibnamefont{Kruerand}},
  \bibinfo{author}{\bibfnamefont{S.~C.} \bibnamefont{Wilks}},
  \bibinfo{author}{\bibfnamefont{J.}~\bibnamefont{Woodworth}},
  \bibinfo{author}{\bibfnamefont{E.~M.} \bibnamefont{Campbell}},
  \bibnamefont{and} \bibinfo{author}{\bibfnamefont{M.~J.} \bibnamefont{Perry}},
  \bibinfo{journal}{Physics of Plasmas} \textbf{\bibinfo{volume}{1}},
  \bibinfo{pages}{1626} (\bibinfo{year}{1994}).

\bibitem[{\citenamefont{Malkin et~al.}(1999)\citenamefont{Malkin, Shvets, and
  Fisch}}]{Fisch3}
\bibinfo{author}{\bibfnamefont{V.~M.} \bibnamefont{Malkin}},
  \bibinfo{author}{\bibfnamefont{G.}~\bibnamefont{Shvets}}, \bibnamefont{and}
  \bibinfo{author}{\bibfnamefont{N.~J.} \bibnamefont{Fisch}},
  \bibinfo{journal}{Phys.~Rev.~Lett.} \textbf{\bibinfo{volume}{82}},
  \bibinfo{pages}{4448} (\bibinfo{year}{1999}).

\bibitem[{\citenamefont{Fisch and Malkin}(1999)}]{Fisch4}
\bibinfo{author}{\bibfnamefont{N.~J.} \bibnamefont{Fisch}} \bibnamefont{and}
  \bibinfo{author}{\bibfnamefont{V.~M.} \bibnamefont{Malkin}},
  \bibinfo{journal}{Phys.~Rev.~Lett.} \textbf{\bibinfo{volume}{82}},
  \bibinfo{pages}{4448} (\bibinfo{year}{1999}).

\bibitem[{\citenamefont{Malkin and Fisch}(2007)}]{Fisch}
\bibinfo{author}{\bibfnamefont{V.~M.} \bibnamefont{Malkin}} \bibnamefont{and}
  \bibinfo{author}{\bibfnamefont{N.~J.} \bibnamefont{Fisch}},
  \bibinfo{journal}{Phys.~Rev.~Lett.} \textbf{\bibinfo{volume}{99}},
  \bibinfo{pages}{205001} (\bibinfo{year}{2007}).

\bibitem[{\citenamefont{Malkin et~al.}(2007)\citenamefont{Malkin, Fisch, and
  Wurtele}}]{Fisch2}
\bibinfo{author}{\bibfnamefont{V.~M.} \bibnamefont{Malkin}},
  \bibinfo{author}{\bibfnamefont{N.~J.} \bibnamefont{Fisch}}, \bibnamefont{and}
  \bibinfo{author}{\bibfnamefont{J.~S.} \bibnamefont{Wurtele}},
  \bibinfo{journal}{Phys.~Rev.~E} \textbf{\bibinfo{volume}{75}},
  \bibinfo{pages}{026404} (\bibinfo{year}{2007}).

\bibitem[{\citenamefont{Ping et~al.}(2005)\citenamefont{Ping, Cheng, Suckewer,
  Clark, Fisch, Hur, and Wurtele}}]{Ping}
\bibinfo{author}{\bibfnamefont{Y.}~\bibnamefont{Ping}},
  \bibinfo{author}{\bibfnamefont{W.}~\bibnamefont{Cheng}},
  \bibinfo{author}{\bibfnamefont{S.}~\bibnamefont{Suckewer}},
  \bibinfo{author}{\bibfnamefont{D.~S.} \bibnamefont{Clark}},
  \bibinfo{author}{\bibfnamefont{N.~J.} \bibnamefont{Fisch}},
  \bibinfo{author}{\bibfnamefont{M.~S.} \bibnamefont{Hur}}, \bibnamefont{and}
  \bibinfo{author}{\bibfnamefont{J.~S.} \bibnamefont{Wurtele}},
  \bibinfo{journal}{Phys.~Rev.~Lett.} \textbf{\bibinfo{volume}{92}},
  \bibinfo{pages}{175007} (\bibinfo{year}{2005}).

\bibitem[{\citenamefont{Cheng et~al.}(2005)\citenamefont{Cheng, Avitzour, Ping,
  Suckewer, Fisch, Hur, and Wurtele}}]{Ping2}
\bibinfo{author}{\bibfnamefont{W.}~\bibnamefont{Cheng}},
  \bibinfo{author}{\bibfnamefont{Y.}~\bibnamefont{Avitzour}},
  \bibinfo{author}{\bibfnamefont{Y.}~\bibnamefont{Ping}},
  \bibinfo{author}{\bibfnamefont{S.}~\bibnamefont{Suckewer}},
  \bibinfo{author}{\bibfnamefont{N.~J.} \bibnamefont{Fisch}},
  \bibinfo{author}{\bibfnamefont{M.~S.} \bibnamefont{Hur}}, \bibnamefont{and}
  \bibinfo{author}{\bibfnamefont{J.~S.} \bibnamefont{Wurtele}},
  \bibinfo{journal}{Phys.~Rev.~Lett.} \textbf{\bibinfo{volume}{94}},
  \bibinfo{pages}{045003} (\bibinfo{year}{2005}).

\bibitem[{\citenamefont{Son and Fisch}(2005)}]{sonprl}
\bibinfo{author}{\bibfnamefont{S.}~\bibnamefont{Son}} \bibnamefont{and}
  \bibinfo{author}{\bibfnamefont{N.~J.} \bibnamefont{Fisch}},
  \bibinfo{journal}{Phys.~Rev.~Lett.} \textbf{\bibinfo{volume}{95}},
  \bibinfo{pages}{225002} (\bibinfo{year}{2005}).

\bibitem[{\citenamefont{Son and Fisch}(2004)}]{sonpla}
\bibinfo{author}{\bibfnamefont{S.}~\bibnamefont{Son}} \bibnamefont{and}
  \bibinfo{author}{\bibfnamefont{N.~J.} \bibnamefont{Fisch}},
  \bibinfo{journal}{Phys.~Lett.~A} \textbf{\bibinfo{volume}{329}},
  \bibinfo{pages}{16} (\bibinfo{year}{2004}).

\bibitem[{\citenamefont{Son and Ku}(2010)}]{IAW}
\bibinfo{author}{\bibfnamefont{S.}~\bibnamefont{Son}} \bibnamefont{and}
  \bibinfo{author}{\bibfnamefont{S.}~\bibnamefont{Ku}},
  \bibinfo{journal}{Phys.~Plasmas} \textbf{\bibinfo{volume}{17}},
  \bibinfo{pages}{024501} (\bibinfo{year}{2010}).

\bibitem[{\citenamefont{Son and Ku}(2009)}]{sonlandau}
\bibinfo{author}{\bibfnamefont{S.}~\bibnamefont{Son}} \bibnamefont{and}
  \bibinfo{author}{\bibfnamefont{S.}~\bibnamefont{Ku}},
  \bibinfo{journal}{Phys.~Plasmas} \textbf{\bibinfo{volume}{17}},
  \bibinfo{pages}{010703} (\bibinfo{year}{2009}).

\bibitem[{\citenamefont{Malkin et~al.}(2000)\citenamefont{Malkin, Tsidulko, and
  Fisch}}]{Fisch5}
\bibinfo{author}{\bibfnamefont{V.~M.} \bibnamefont{Malkin}},
  \bibinfo{author}{\bibfnamefont{Y.}~\bibnamefont{Tsidulko}}, \bibnamefont{and}
  \bibinfo{author}{\bibfnamefont{N.~J.} \bibnamefont{Fisch}},
  \bibinfo{journal}{Phys.~Rev.~Lett.} \textbf{\bibinfo{volume}{85}},
  \bibinfo{pages}{4068} (\bibinfo{year}{2000}).

\bibitem[{\citenamefont{Sturm}(1976)}]{Sturm}
\bibinfo{author}{\bibfnamefont{K.}~\bibnamefont{Sturm}},
  \bibinfo{journal}{Z.~Physik~B} \textbf{\bibinfo{volume}{25}},
  \bibinfo{pages}{247} (\bibinfo{year}{1976}).

\bibitem[{\citenamefont{Sturm}(1977)}]{Sturm2}
\bibinfo{author}{\bibfnamefont{K.}~\bibnamefont{Sturm}}, \bibinfo{journal}{Z.
  Physik B} \textbf{\bibinfo{volume}{28}}, \bibinfo{pages}{1}
  (\bibinfo{year}{1977}).

\bibitem[{\citenamefont{Gibbons et~al.}(1976)\citenamefont{Gibbons,
  Schnatterly, Ritsko, and Fields}}]{Gibbons}
\bibinfo{author}{\bibfnamefont{P.~C.} \bibnamefont{Gibbons}},
  \bibinfo{author}{\bibfnamefont{S.~E.} \bibnamefont{Schnatterly}},
  \bibinfo{author}{\bibfnamefont{J.~J.} \bibnamefont{Ritsko}},
  \bibnamefont{and} \bibinfo{author}{\bibfnamefont{J.~R.}
  \bibnamefont{Fields}}, \bibinfo{journal}{Phys.~Rev.~B}
  \textbf{\bibinfo{volume}{13}}, \bibinfo{pages}{2451} (\bibinfo{year}{1976}).

\bibitem[{\citenamefont{Ku and Eguiluz}(1999)}]{Ku}
\bibinfo{author}{\bibfnamefont{W.}~\bibnamefont{Ku}} \bibnamefont{and}
  \bibinfo{author}{\bibfnamefont{A.~G.} \bibnamefont{Eguiluz}},
  \bibinfo{journal}{Phys.~Rev.~Lett.} \textbf{\bibinfo{volume}{82}},
  \bibinfo{pages}{2350} (\bibinfo{year}{1999}).

\bibitem[{\citenamefont{Ku and Eguiluz}(2000)}]{Ku2}
\bibinfo{author}{\bibfnamefont{W.}~\bibnamefont{Ku}} \bibnamefont{and}
  \bibinfo{author}{\bibfnamefont{A.~G.} \bibnamefont{Eguiluz}},
  \bibinfo{journal}{J.~Phys.~Chem.~Solids} \textbf{\bibinfo{volume}{61}},
  \bibinfo{pages}{383} (\bibinfo{year}{2000}).

\bibitem[{\citenamefont{Pitarke and Campillo}(2000)}]{Pitarke}
\bibinfo{author}{\bibfnamefont{J.~M.} \bibnamefont{Pitarke}} \bibnamefont{and}
  \bibinfo{author}{\bibfnamefont{I.}~\bibnamefont{Campillo}},
  \bibinfo{journal}{J.~Phys.~Chem.~Solids} \textbf{\bibinfo{volume}{164}},
  \bibinfo{pages}{147} (\bibinfo{year}{2000}).

\bibitem[{\citenamefont{Tirao et~al.}(2007)\citenamefont{Tirao, Stutz, Silkin,
  Chulkov, and Cusatis}}]{Thira}
\bibinfo{author}{\bibfnamefont{G.}~\bibnamefont{Tirao}},
  \bibinfo{author}{\bibfnamefont{G.}~\bibnamefont{Stutz}},
  \bibinfo{author}{\bibfnamefont{V.~M.} \bibnamefont{Silkin}},
  \bibinfo{author}{\bibfnamefont{E.~V.} \bibnamefont{Chulkov}},
  \bibnamefont{and} \bibinfo{author}{\bibfnamefont{C.}~\bibnamefont{Cusatis}},
  \bibinfo{journal}{J.~ phys.: Condens.~Matter} \textbf{\bibinfo{volume}{19}},
  \bibinfo{pages}{046207} (\bibinfo{year}{2007}).

\bibitem[{\citenamefont{McKinstrie and Simon}(1986)}]{McKinstrie}
\bibinfo{author}{\bibfnamefont{C.~J.} \bibnamefont{McKinstrie}}
  \bibnamefont{and} \bibinfo{author}{\bibfnamefont{A.}~\bibnamefont{Simon}},
  \bibinfo{journal}{Phys.~Fluids} \textbf{\bibinfo{volume}{29}},
  \bibinfo{pages}{1959} (\bibinfo{year}{1986}).

\bibitem[{\citenamefont{Lindhard}(1954)}]{Lindhard}
\bibinfo{author}{\bibfnamefont{J.}~\bibnamefont{Lindhard}},
  \bibinfo{journal}{K. Dan. Vidensk. Sels. Mat. Fys. Medd}
  \textbf{\bibinfo{volume}{28}}, \bibinfo{pages}{8} (\bibinfo{year}{1954}).

\bibitem[{\citenamefont{Capjack et~al.}(1982)\citenamefont{Capjack, James, and
  McMullin}}]{Capjack}
\bibinfo{author}{\bibfnamefont{C.~E.} \bibnamefont{Capjack}},
  \bibinfo{author}{\bibfnamefont{C.~R.} \bibnamefont{James}}, \bibnamefont{and}
  \bibinfo{author}{\bibfnamefont{J.~N.~.} \bibnamefont{McMullin}},
  \bibinfo{journal}{J.~Appl.~Phys.} \textbf{\bibinfo{volume}{53}},
  \bibinfo{pages}{4046} (\bibinfo{year}{1982}).

\bibitem[{\citenamefont{DuBois and Kivelson}(1969)}]{DuBois}
\bibinfo{author}{\bibfnamefont{D.~F.} \bibnamefont{DuBois}} \bibnamefont{and}
  \bibinfo{author}{\bibfnamefont{M.~C.} \bibnamefont{Kivelson}},
  \bibinfo{journal}{Phys. Rev.} \textbf{\bibinfo{volume}{186}},
  \bibinfo{pages}{409} (\bibinfo{year}{1969}).

\bibitem[{\citenamefont{Zacharias}(1975)}]{Zacharias}
\bibinfo{author}{\bibfnamefont{P.~J.} \bibnamefont{Zacharias}},
  \bibinfo{journal}{J.~Phys.~ F: Metal Phys.} \textbf{\bibinfo{volume}{5}},
  \bibinfo{pages}{645} (\bibinfo{year}{1975}).

\bibitem[{\citenamefont{Urner-Wille and Raether}(1976)}]{Urner-Wille}
\bibinfo{author}{\bibfnamefont{M.}~\bibnamefont{Urner-Wille}} \bibnamefont{and}
  \bibinfo{author}{\bibfnamefont{H.}~\bibnamefont{Raether}},
  \bibinfo{journal}{Phys. Lett.} \textbf{\bibinfo{volume}{58A}},
  \bibinfo{pages}{265} (\bibinfo{year}{1976}).

\bibitem[{\citenamefont{Kloos}(1973)}]{Kloos}
\bibinfo{author}{\bibfnamefont{T.}~\bibnamefont{Kloos}},
  \bibinfo{journal}{Z.~Physik} \textbf{\bibinfo{volume}{265}},
  \bibinfo{pages}{225} (\bibinfo{year}{1973}).

\bibitem[{\citenamefont{Son and Fisch}(2006)}]{son2}
\bibinfo{author}{\bibfnamefont{S.}~\bibnamefont{Son}} \bibnamefont{and}
  \bibinfo{author}{\bibfnamefont{N.~J.} \bibnamefont{Fisch}},
  \bibinfo{journal}{Phys.~Lett.~A} \textbf{\bibinfo{volume}{356}},
  \bibinfo{pages}{72} (\bibinfo{year}{2006}).

\bibitem[{\citenamefont{Drake and Batha}(1991)}]{Drake}
\bibinfo{author}{\bibfnamefont{R.~P.} \bibnamefont{Drake}} \bibnamefont{and}
  \bibinfo{author}{\bibfnamefont{S.~H.} \bibnamefont{Batha}},
  \bibinfo{journal}{Phys.~Fluids B} \textbf{\bibinfo{volume}{3}},
  \bibinfo{pages}{2936} (\bibinfo{year}{1991}).

\bibitem[{\citenamefont{Rose and Russel}(2001)}]{Rose}
\bibinfo{author}{\bibfnamefont{H.}~\bibnamefont{Rose}} \bibnamefont{and}
  \bibinfo{author}{\bibfnamefont{D.~A.} \bibnamefont{Russel}},
  \bibinfo{journal}{Physics of Plasmas} \textbf{\bibinfo{volume}{8}},
  \bibinfo{pages}{4784} (\bibinfo{year}{2001}).

\bibitem[{\citenamefont{Morales and O'Neil}(1972)}]{Oneil}
\bibinfo{author}{\bibfnamefont{G.~J.} \bibnamefont{Morales}} \bibnamefont{and}
  \bibinfo{author}{\bibfnamefont{T.~M.} \bibnamefont{O'Neil}},
  \bibinfo{journal}{Phys.~Rev.~Lett.} \textbf{\bibinfo{volume}{28}},
  \bibinfo{pages}{417} (\bibinfo{year}{1972}).

\end{thebibliography}

\end{document}